# Digital Voices of Survival: From Social Media Disclosures to Support Provisions for Domestic Violence Victims

Kanlun Wang
Fairfield University
kanlun.wang@fairfield.edu

Zhe Fu
UNC Charlotte
zfu2@charlotte.edu

Wangjiaxuan Xin
UNC Charlotte
wxin@charlotte.edu

Shashi Kiran Chandrappa
Fairfield University
schandrappa@student.fairfield.edu

Lina Zhou
UNC Charlotte
lzhou8@charlotte.edu

## Abstract

*Domestic Violence (DV) is a pervasive public health problem characterized by patterns of coercive and abusive behavior within intimate relationships. With the rise of social media as a key outlet for DV victims to disclose their experiences, online self-disclosure has emerged as a critical yet underexplored avenue for support-seeking. In addition, existing research lacks a comprehensive and nuanced understanding of DV self-disclosure, support provisions, and their connections. To address these gaps, this study proposes a novel computational framework for modeling DV support-seeking behavior alongside community support mechanisms. The framework consists of four key components: self-disclosure detection, post clustering, topic summarization, and support extraction and mapping. We implement and evaluate the framework with data collected from relevant social media communities. Our findings not only advance existing knowledge on DV self-disclosure and online support provisions but also enable victim-centered digital interventions.*

**Keywords:** domestic violence, support provisions, classification, clustering, large language models.

## 1. Introduction

Domestic Violence (DV) represents a significant social and public health concern characterized by systematic patterns of abusive behaviors, ranging from physical and sexual aggression to emotional, psychological, economic, and increasingly, technological forms of coercion (Aldkheel et al., 2021; Christia et al., 2023). This form of violence cuts across diverse cultural, economic, and demographic groups globally, resulting in substantial negative impacts on victims' physical health, emotional well-being, and long-term personal development (Buzawa & Buzawa, 2003). Consequently, the severity and prevalence of DV have prompted the development of intervention measures, prevention strategies, and informed policies to mitigate its effects.

With the rise of digital platforms, social media has become a pivotal environment where DV victims increasingly share their personal experiences, signaling a notable departure from traditional patterns of support-seeking behaviors (Aldkheel et al., 2021; Gundersen & Zaleski, 2021; Zhang & Li, 2025). Self-disclosure, in the context of DV research, refers to the personal and voluntary act of revealing one's own past or ongoing experiences of DV within a communication setting (Aldkheel et al., 2021; Christia et al., 2023). In contrast, non-self-disclosure refers to online DV-related content that does not involve first-person revelation of one's own experiences. For instance, it may include third-person accounts (e.g., sharing news stories), general opinions, advocacy or awareness posts, informational resources, or discussions about DV without personal victimization being disclosed. Online self-disclosure of DV experiences represents a transformative shift that capitalizes on the unique affordances of digital platforms, such as anonymity, ease of access, and rapid dissemination, to enable victims to disclose abuse experiences, seek support, and engage with broad communities. These disclosures not only help victims alleviate feelings of isolation but also provide access to emotional and informational support. Furthermore, it also plays a critical role in increasing societal awareness, influencing public opinions, and potentially shaping policy decision-making regarding DV and victim assistance services (Christia et al., 2023; Gueta et al., 2024; McCauley et al., 2018).

However, accurate identification of self-disclosure posts related to DV in social media remains a significant methodological challenge. This complexity stems from the nuanced and often implicit language used by victims, the variability in how abuse is described, and the difficulty of distinguishing personal disclosures from general discussions or informational content. There are very few related studies, which have focused on developing theoretical frameworks for classifying DV dynamics (e.g., Holtzworth-Munroe & Meehan, 2004; Johnson, 2008) and computational models leveraging machine learning techniques for detecting DV intent (Subramani et al., 2017) and predicting recidivism (Wijenayake et al., 2018).

Moreover, understanding the support-seeking behaviors of DV victims is a critical area of inquiry with significant implications for the development of effective support systems, evidence-based policy, and targeted interventions. A growing body of literature (e.g., Huntley et al., 2019; Ruiz-Pérez et al., 2007) has documented the multifaceted barriers that hinder victims from accessing timely assistance, including fear of retaliation, social stigma, limited resources, and structural or cultural constraints. Investigating the pathways through which victims seek support, across both formal services and informal networks, offers valuable insight into unmet needs and informs the design of more tailored, victim-centered supports (Wells et al., 2024).

Despite growing scholarly attention to DV, several critical gaps remain in understanding the diffusion of DV experience sharing and support provision in online communities. First, prior research has largely overlooked the specific challenge of distinguishing DV-related self-disclosures from non-self-disclosures in digital contexts. This leaves a significant gap in developing automated and scalable approaches to detect and classify DV-related self-disclosures across platforms. Second, while the importance of victim-centered support and intervention mechanisms is widely recognized, there is a lack of a comprehensive framework that systematically captures the distinct types of DV represented in online self-disclosures and aligns them with the appropriate forms of community support.

To address the aforementioned research gaps, this study seeks to answer the following questions and develop a comprehensive framework that provides tailored support(s) to DV victims under various types of abusive relationships:

(1) How effectively can self-disclosure content be differentiated from non-self-disclosure content on social media platforms?

(2) How can distinct types of topics in DV self-disclosure content be systematically identified and characterized?

(3) What are the different categories of support provisions for DV victims? How do they relate to DV self-disclosure topics?

In line with Gregor & Hevner's (2013) Design Science Research framework, this study advances both theory and practice by developing and evaluating a novel computational framework for online DV support-seeking. The artifact integrates an LLM-based classifier to detect DV self-disclosure, unsupervised clustering to uncover thematic patterns in victim narratives, and mapping techniques to align these clusters with community-provided support provisions. Theoretically, the study contributes by clarifying the distinction between self-disclosure and non-self-disclosure in online contexts and systematically linking disclosure themes to appropriate forms of support. Practically, it offers a scalable, victim-centered approach that can inform the design of automated tools for triaging victims' needs and guiding digital interventions across social media, healthcare, and community platforms.

The rest of this paper is organized as follows. Section 2 reviews related work on DV self-disclosure, classification, and online help-seeking. Section 3 introduces our proposed framework for DV support-seeking, and Section 4 explains the evaluation design. Section 5 presents the results, and Section 6 concludes the paper along with future research directions.

## 2. Related Work

This section reviews three areas of related literature on DV: online self-disclosure of DV experiences, classification of DV-related content, and support-seeking behaviors of DV victims.

### 2.1. Online self-disclosure of DV experiences

Social media has become an essential outlet for victims to disclose their experiences. The anonymity, immediacy, and broad reach offered by these platforms have lowered barriers to disclosure, particularly for experiences that are often stigmatized or silenced in traditional offline contexts. These digital environments not only provide a safe space for victims to articulate their trauma but also influence societal perceptions of abuse. Notably, online campaigns such as #MaybeHeDoesntHitYou (McCauley et al., 2018) have raised the awareness of non-physical forms of abuse, including emotional and psychological violence (Aldkheel et al., 2021; Christia et al., 2023; Gueta et al., 2024), thereby reshaping the public discourse on DV.

Multiple studies have identified key motivations driving DV victims to disclose their experiences online. These include the pursuit of emotional and informational support (Aldkheel et al., 2021), the need for validation (Zhang & Li, 2025), the desire to find community among others with shared experiences (Gundersen & Zaleski, 2021), and the potential therapeutic effects of narrative expression (Murvartian et al., 2024). Online self-disclosure also empowers victims to reclaim agency over their personal narratives and contributes to broader social awareness through viral movements such as #MeToo (Gorissen et al., 2023). Importantly, online disclosures may also serve as a conduit to critical resources, including legal aid, mental health services, and safety planning tools (Christia et al., 2023; Gueta et al., 2024; McCauley et al., 2018).

## 2.2. DV classification

DV constitutes a complex and multifaceted pattern of abusive behaviors, encompassing physical, sexual, emotional, psychological, economic, and technological forms of abuse (Buzawa & Buzawa, 2003).

The literature offers several typologies for systematically categorizing DV. For instance, Johnson's widely cited framework (Johnson, 2008) distinguishes the forms of intimate partner violence based on their underlying dynamics of power and control. Specifically, the framework identifies "situational couple violence," characterized by episodic and often mutual aggression triggered by specific conflicts, and "intimate terrorism," which is typified by chronic, coercively controlling behaviors predominantly exerted by one partner. Complementarily, the typology developed by Holtzworth-Munroe & Meehan (2004) classifies perpetrators based on their psychological profiles and behavior patterns, delineating categories such as family-only offenders who confine their abuse strictly to intimate relationships, dysphoric-borderline offenders who exhibit emotional instability and intense jealousy, and so on.

Beyond traditional DV typologies, computational methods, particularly machine learning, are gaining recognition in DV research. Hui et al. (2023) highlights how machine learning techniques provide new avenues for addressing DV. Recent work on intent detection from social media communications uses machine learning techniques and Natural Language Processing (NLP) to identify subtle cues of victims' emotional states, distress signals, or explicit support-seeking behaviors (Subramani et al., 2017). Additionally, multi-class classification techniques have been used to categorize DV-related content into personal disclosures, informational posts, and advocacy or awareness-related messages (Subramani et al., 2019). Predictive modeling techniques, such as decision trees, random forests, and neural networks, have been used to assess offender recidivism risk, aiding judicial and rehabilitative interventions (Wijenayake et al., 2018). Lastly, Latent Dirichlet Allocation (LDA) is employed to identify DV-related topics, and Term Frequency–Inverse Document Frequency-based feature engineering is applied to support multi-class classification of DV self-disclosure content using a range of traditional machine learning algorithms (Salehi & Ghahari, 2024).

## 2.3. DV support-seeking behaviors

Victims' support-seeking behaviors in cases of DV represent a complex and nuanced phenomenon, deeply influenced by an intricate interplay of individual, relational, and broader sociocultural factors. Research consistently reveals distinct support-seeking behaviors and experiences between male and female DV victims (Huntley et al., 2019; Wells et al., 2024). In addition, victim support-seeking strategies also exhibit substantial variation, ranging from proactive engagement with formal resources like shelters and helplines to reliance on informal support networks such as family and friends, or minimal engagement altogether (Ben-Porat, 2020). This variability is well-documented among college populations experiencing sexual assault, dating violence, and stalking, where distinct patterns emerge based on factors such as perceived severity, availability of support, and societal attitudes toward victimization (Ameral et al., 2020).

Despite these varying approaches, significant and consistent barriers impede victims' willingness or ability to seek assistance. The inherently private nature of DV fosters silence and concealment, driven by feelings of fear, guilt, shame, and societal taboos (Ruiz-Pérez et al., 2007). Victims frequently normalize their experiences or fear judgment and retribution, leading to underreporting and avoidance of formal support channels (Fanslow & Robinson, 2010). Prominent obstacles identified in the literature include fear of further violence, anxiety over potential child custody issues, and insufficient familial support (Fanslow & Robinson, 2010). Healthcare settings, despite being crucial intervention points, are often ill-equipped to respond effectively due to inadequate training, limited resources, time constraints, and insufficient confidential spaces (Purbarrar et al., 2023). Furthermore, prior negative experiences with law enforcement or criminal justice systems can significantly reduce victims' likelihood of engaging these services again in the future (Kunst et al., 2015).

Although existing research provides valuable DV typologies and computational methods for classifying related content, it does not adequately distinguish between self-disclosure and non-self-disclosure in online environments. Moreover, studies on support-seeking behaviors lack a structured framework that connects specific types of disclosure and the appropriate support required. These gaps lead to a fragmented understanding and create a significant risk of mismatching victims' needs with available resources.

## 3. A Framework of DV Support-Seeking

The objective of this research is to create a comprehensive framework that integrates state-of-the-art Large Language Models (LLMs) with advanced clustering techniques for the automated detection of DV self-disclosure and identification of the themes of associated support(s). As illustrated in Figure 1, the proposed framework is built upon four key components: (1) LLM-based self-disclosure detection, (2) post clustering, (3) topic summarization, and (4) support extraction and mapping.

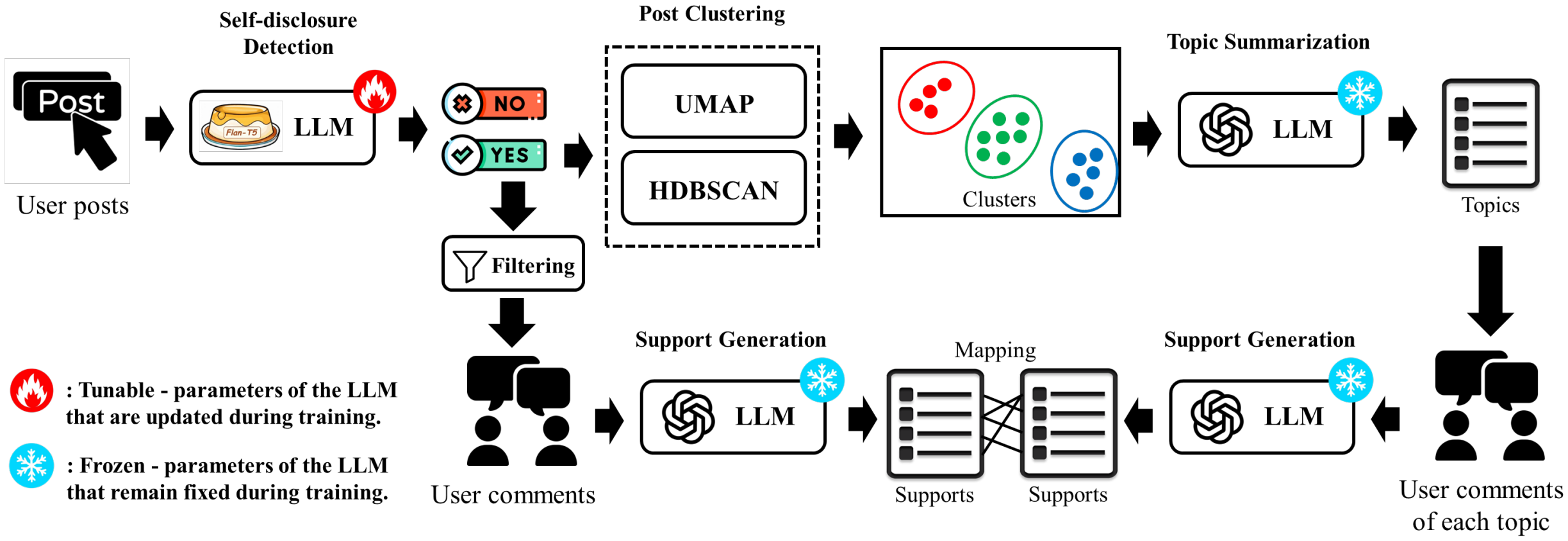

**Figure 1. A computational framework for modeling DV support-seeking behaviors and support provisions**

### 3.1 LLM-based self-disclosure detection

Given a set of social media posts, this component aims to predict whether each post contains a self-disclosure of DV. We formulate DV self-disclosure as a binary classification task.

Let $c$ denote the textual content of a social media post, containing a post title and body. We focus on training a classifier $f(\cdot)$ for DV self-disclosure (see Equation 1):

$$y = f(c, \Theta) \qquad (1)$$

where $y$ denotes a classification result for the target post $c$, and $\Theta$ denotes the set of parameters of $f(\cdot)$.

Training classification models requires labelled data. The recent surge of LLMs brings an unprecedented opportunity to automate the ground truth labeling process. Several studies (e.g., He et al., 2024; Tan et al., 2024) have demonstrated that LLMs, with sufficient guidance, can achieve annotation quality comparable to or even better than that of a human standard. Therefore, we leverage an LLM for post annotation with self-disclosure labels.

To be specific, we conduct instruction tuning on Flan-T5 (the 11B version) (Chung et al., 2024), an open-source LLM released by Google. As to the instruction-formatted training samples, we leverage 700 instruction-formatted training samples (equally split between self-disclosure content ("Yes" class) and non-self-disclosure content ("No" class)) from our human-annotated data. For each training sample, we design the prompt for the DV annotation task based on the template provided by Alpaca (Taori et al., 2023) and append task instructions that introduce the concept of DV, along with an input query and an output (see Table 1).

Through these instruction-formatted training samples, we convert a language generation task into a binary classification task by instructing the LLM to respond with a binary answer of "Yes" or "No".

**Table1. A prompt template for DV annotation**

| **LLM Input** | |
|---|---|
| *Task Instruction* | A self-disclosure of domestic violence refers to the personal and voluntary act of sharing one's own experiences as a victim with another individual, a group, or the general public. |
| *Input Query* | Given a social media post: [*Post Content*], classify whether it is a self-disclosure of domestic violence or not by answering 'Yes' or 'No'. Answer: |
| **LLM Output** | |
| *Output* | *"Yes" or "No"* |

## 3.2. Post clustering

In view of the diversity of DV-related social media posts, we first group the posts based on their semantic distance to uncover meaningful topic patterns. To this end, we adopt the Hierarchical Density-Based Spatial Clustering of Applications with Noise algorithm (HDBSCAN) (Malzer & Baum, 2020), a density-based clustering algorithm using cluster stability as the quality measure. It is particularly well-suited for this task due to several advantages over traditional methods, such as K-means or hierarchical clustering. Specifically, HDBSCAN does not require prespecifying the number of clusters, but instead can identify clusters of varying density, and automatically classify noise or outliers, which is valuable when dealing with noisy and unstructured social media data.

When encoding the DV social media content for clustering, the widely-adopted transformer-based pre-trained language models often suffer from the issue of non-smooth anisotropic distribution (Huang et al., 2021). The dimensions in such generated word embeddings do not all contribute to the content semantics, partly because some dimensions may represent the structure or position information. As a result, those word embeddings are not suitable for content semantic distance search in clustering algorithms. To address this issue and enhance the effectiveness of HDBSCAN, we employ Sentence-BERT (Reimers & Gurevych, 2019) to encode the semantic content of each post into dense, contextualized sentence embeddings optimized for similarity tasks. Since the resulting embedding space is high-dimensional, we apply Uniform Manifold Approximation and Projection (UMAP), a non-linear dimensionality reduction technique, to project the embeddings into a lower-dimensional space that preserves local semantic relationships while reducing noise and redundancy. This reduced space serves as the input for HDBSCAN to generate coherent and interpretable topic clusters.

## 3.3 Topic summarization

To enhance the interpretability of clustered social media posts, we perform topic summarization for each cluster to identify the underlying DV-related topics. This step is essential for translating unsupervised clustering results into meaningful insights that can inform intervention strategies, support services, and policy development.

In comparison to traditional topic modeling techniques, such as LDA and BERTopic, LLMs offer significant advantages: they generate more semantically coherent topics, do not rely on predefined word distributions, and are better equipped to capture nuanced contextual meanings that extend beyond surface-level word co-occurrences (Yang & Kim, 2025).

To be specific, we first combine all the posts within each cluster as a document and then generate topics from this document using the GPT-4o model (OpenAI et al., 2024), which is an open-weight reasoning LLM with state-of-the-art performances across various NLP tasks. In addition, we use a well-designed prompt guided by Alpaca (Taori et al., 2023) for the GPT-4o model, which consists of two components: a detailed task instruction and an input query. The task instruction is used to provide a general outline of the task requirements, and the input query contains the current input data for the task, as illustrated in Table 2.

**Table 2. A prompt template for topic summarization within each cluster**

| LLM Input | |
|---|---|
| *Task Instruction* | Given a set of posts grouped into *N* (based on the outputs of DBSCAN) distinct clusters, all related to domestic violence, generate a concise summary for each cluster that captures its main theme(s) or topic(s). |
| *Input Query* | Here are *N* clusters of posts: [*Post Content*]. Generate a concise summary for each cluster that captures its main topic(s). Answer: |
| **LLM Output** | |
| *Output* | [*A list of topic(s) for each cluster*] |

**Table 3. A prompt template for general support summarization**

| LLM Input | |
|---|---|
| *Task Instruction* | Given a set of user comments on domestic violence self-disclosure (posts) that seek support, extract and summarize the forms of supports provided to the victims from the comments. Ensure that the forms of supports are distinct from one another. |
| *Input Query* | Here are the comments on domestic violence self-disclosure that seek support: [*Top-10 Comment Content across All Clusters*]. Extract and summarize the forms of supports provided to the victims from the comments. |
| **LLM Output** | |
| *Output* | [*A list of supports for domestic violence victims*] |

**Table 4. A prompt template for support summarization from each cluster**

| LLM Input | |
|---|---|
| *Task Instruction* | Given a set of user comments on domestic violence self-disclosure (posts) that seek support, extract and summarize the forms of supports provided to the victims from the comments. |
| *Input Query* | Here are the comments on domestic violence self-disclosure that seek support: [*Top-10 Comment Content within Each Cluster*]. Extract and summarize the forms of supports provided to the victims from the comments. |
| **LLM Output** | |
| *Output* | [*A list of supports for domestic violence victims*] |

## 3.4 Support extraction and mapping

To identify the support provided by community members in response to the summarized DV topics, we develop a novel method to extract informative and supportive content from user comments on the posts grouped within individual clusters.

First, we construct a pool of candidate forms of support by summarizing user-generated comments on the DV self-disclosure posts across all clusters using an LLM. Like topic summarization, we adopt GPT-4o (OpenAI et al., 2024) for the support extraction task by employing an instruction-based prompt (see Table 3) guided by Alpaca (Taori et al., 2023). However, the total length of user comments associated with DV posts exceeds GPT-4o's maximum input limit of 128,000 tokens. To address the issue, we apply a filtering strategy based on the karma scores of comments by selecting the top 10 user comments with the highest karma scores from each cluster to serve as input for the LLM. This approach ensures adherence to the model's input constraints while also prioritizing the most informative and contextually meaningful content.

In addition, we extract representative support types from user comments on DV self-disclosure posts within each cluster by employing GPT-4o (OpenAI et al., 2024) with an instruction-based prompt informed by Alpaca (Taori et al., 2023), as presented in Table 4. We then analyze the relationships between the supports generated across clusters and those identified within individual clusters, mapping them back to the DV topics outlined in Section 3.3.

# 4. Evaluation

## 4.1. Data collection and pre-processing

Reddit is chosen as the primary data source because it is a widely used social media platform that allows users to share, discuss, and rate content. Its anonymous nature encourages users to openly share personal and sensitive experiences, such as those involving DV.

Reddit is organized into subreddits—communities focused on specific topics—many of which are directly or indirectly related to DV. To identify relevant communities, we conduct a keyword-based search using the terms "domestic violence" and "abusive." Three researchers independently review the resulting subreddits by examining their descriptions and sampling post content. Through this vetting process, twelve subreddits are identified as thematically aligned with DV-related discourse: r/domesticviolence, relationships, AbuseInterrupted, abusiverelationships, traumatoolbox, familycourt, abusiveparents, raisedbynarcissists, abusivesiblings, insaneparents, relationship_advice, and emotionalabuse.

We employ the Reddit official API (PRAW) to collect all available historical posts from the selected subreddits. Since this study focuses on how users seek and receive support regarding their DV experiences, we retain only posts that have received at least one comment. This filtering criterion ensures that the data reflect interactive engagement, enhancing the ecological validity of our findings. This step results in a curated dataset of 9,013 posts. Each post includes metadata such as the post title, body text, timestamp, and associated comments.

To enable supervised learning, we create a labeled subset of the data that is randomly selected from the initial dataset. One researcher manually annotates 700 randomly selected posts, assigning binary labels to indicate whether each post constitutes a DV self-disclosure or not. Another two researchers review the annotations for consistency. All inconsistent labels are resolved through discussion among the three researchers. The labeled subset is constructed to ensure class balance, containing an equal number of self-disclosure and non-self-disclosure posts. The remaining 8,313 unlabeled posts are reserved for the classification task following model validation and robustness checks.

## 4.2. Model evaluation

To evaluate the performance of the classification model for DV self-disclosure, we use standard metrics: accuracy, precision, recall, and F1 score. We also apply 10-fold cross-validation, where the dataset is split into 10 parts, with nine used for training and the remaining one for testing. The learning rate is set to $3 \times 10^{-5}$.

For the evaluation of topic summarization and support extraction, we set the minimum cluster size to 20 and require at least 20 samples per cluster. Additionally, we use Sentence-BERT embeddings with a dimensionality of 384.

## 5. Results

This section reports the results for DV self-disclosure detection as well as extracting DV topics along with their associated support categories based on clustering analysis results.

### 5.1. DV self-disclosure detection

Figure 2 presents the performance of the DV self-disclosure classification in each fold, where the x-axis denotes the fold number. The model achieves an average accuracy of 82.01% (±4.80%), precision of 82.01% (±5.58%), recall of 82.57% (±9.38%), and an F1 score of 81.94% (±5.41%). The model demonstrates both effectiveness and robustness in detecting DV self-disclosure. Thus, we proceed with classifying the remaining unlabeled data using the trained model. The model classifies 2,712 posts as self-disclosure and 5,601 as non-self-disclosure. After merging the classified data with the manually annotated subset, the final dataset consists of a total of 3,062 self-disclosure and 5,951 non-self-disclosure posts.

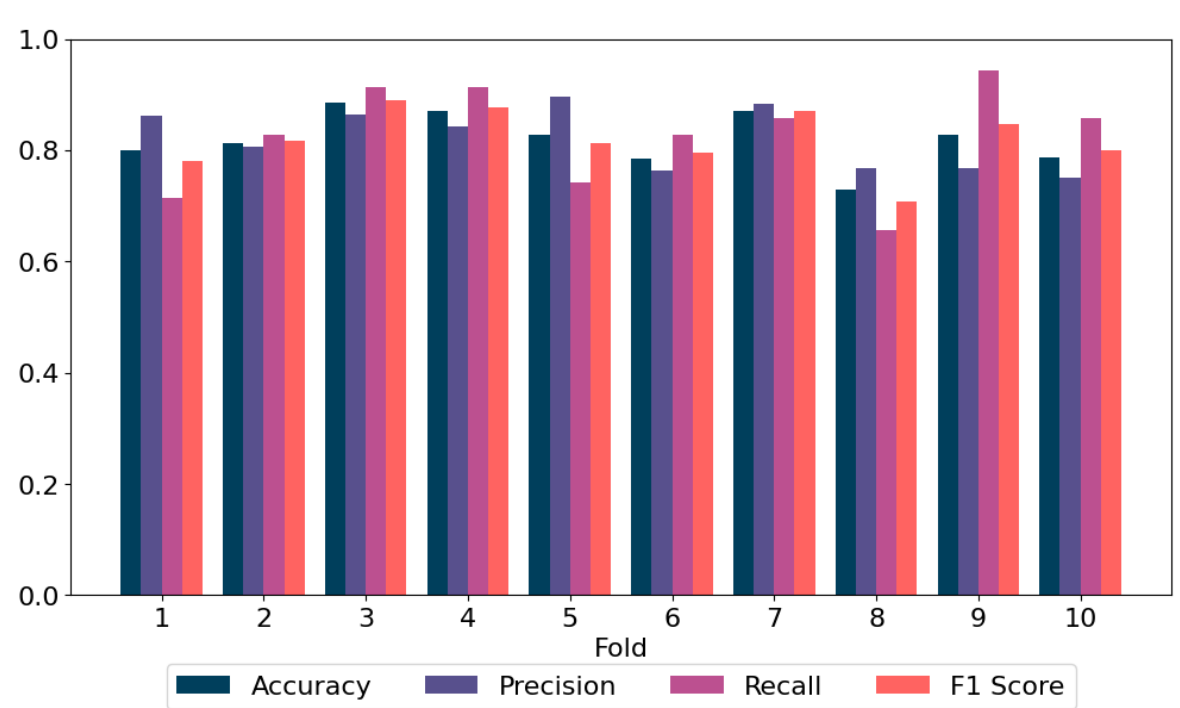

**Figure 2. DV Self-disclosure detection performance**

### 5.2. Clusters/topics of self-disclosures

The clustering results of DV self-disclosure are presented in Figure 3, and the summarized DV topics across all clusters are presented on the left side of Figure 4. The figures show 20 thematically coherent groups of DV-related topics, with a few large clusters standing out both in size and semantic depth. For instance, cluster 0, the most populous one, centers on emotional turmoil in intimate relationships, capturing narratives of psychological manipulation, deep emotional confusion, and suffering in romantic contexts. Cluster 2, which emerged as another substantial grouping, is dominated by posts on custody and legal battles. These narratives detail traumatic encounters with family courts and abusive ex-partners, revealing both institutional barriers and the protracted nature of post-separation abuse. Cluster 8 focuses on the dynamics of abuse and victimization, offering granular descriptions of abusive patterns, power imbalances, and victims' understanding of control mechanisms. Similarly, Cluster 12 highlights law enforcement involvement, where victims recount their interactions with police, often reflecting either inadequate support or procedural failures. Interestingly, Cluster 14 captures internal conflict and self-doubt, illustrating how many victims wrestle with second-guessing their perceptions and experience identity dissonance even after recognizing the abuse. Finally, Cluster 19 represents love, loss, and emotional damage, where emotional aftermath, heartbreak, and betrayal surface as major themes.

The remaining smaller clusters, though less dense, enrich the thematic landscape. These include discussions of gaslighting (Cluster 11), pregnancy and motherhood during abuse (Cluster 15), academic perspectives on trauma (Cluster 3), and narcissistic abuse (Cluster 6), among others.

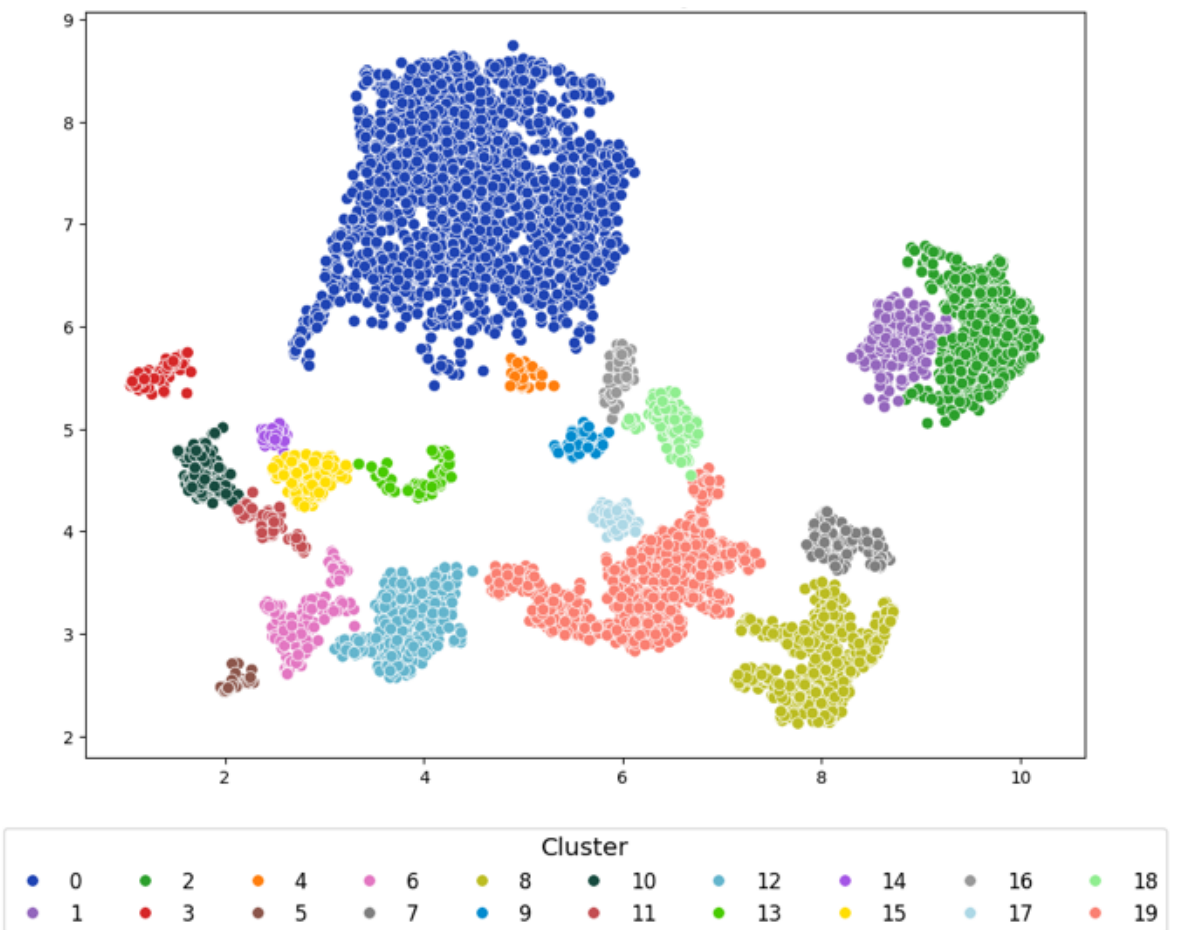

**Figure 3. DV self-disclosure clustering results**

### 5.3. Support provisions

The extracted DV self-disclosure topics and their corresponding online support provided are shown on the right side of Figure 4. Particularly, support 1 (prioritize safety and healing) appears most frequently, across 11 topic clusters, especially in DV posts about emotional abuse, disbelief, and recovery. Support 3 (seek professional support) is also widely used across different clusters involving emotional manipulation, long-term trauma, and gaslighting, highlighting the need for therapy and guidance. Support 4 (build a support system) and Support 13 (empower through knowledge) often appear in clusters about education, recovery, and parenting, emphasizing the value of both knowledge and information. Support 10 (let go of shame) is provided in posts about legal struggles, healing, and law enforcement, showing how social stigma or shame can be a barrier to seeking support. Support 14 (assert legal and parental rights) is common in legal- and custody-related posts, encouraging victims to recognize and exercise their rights.

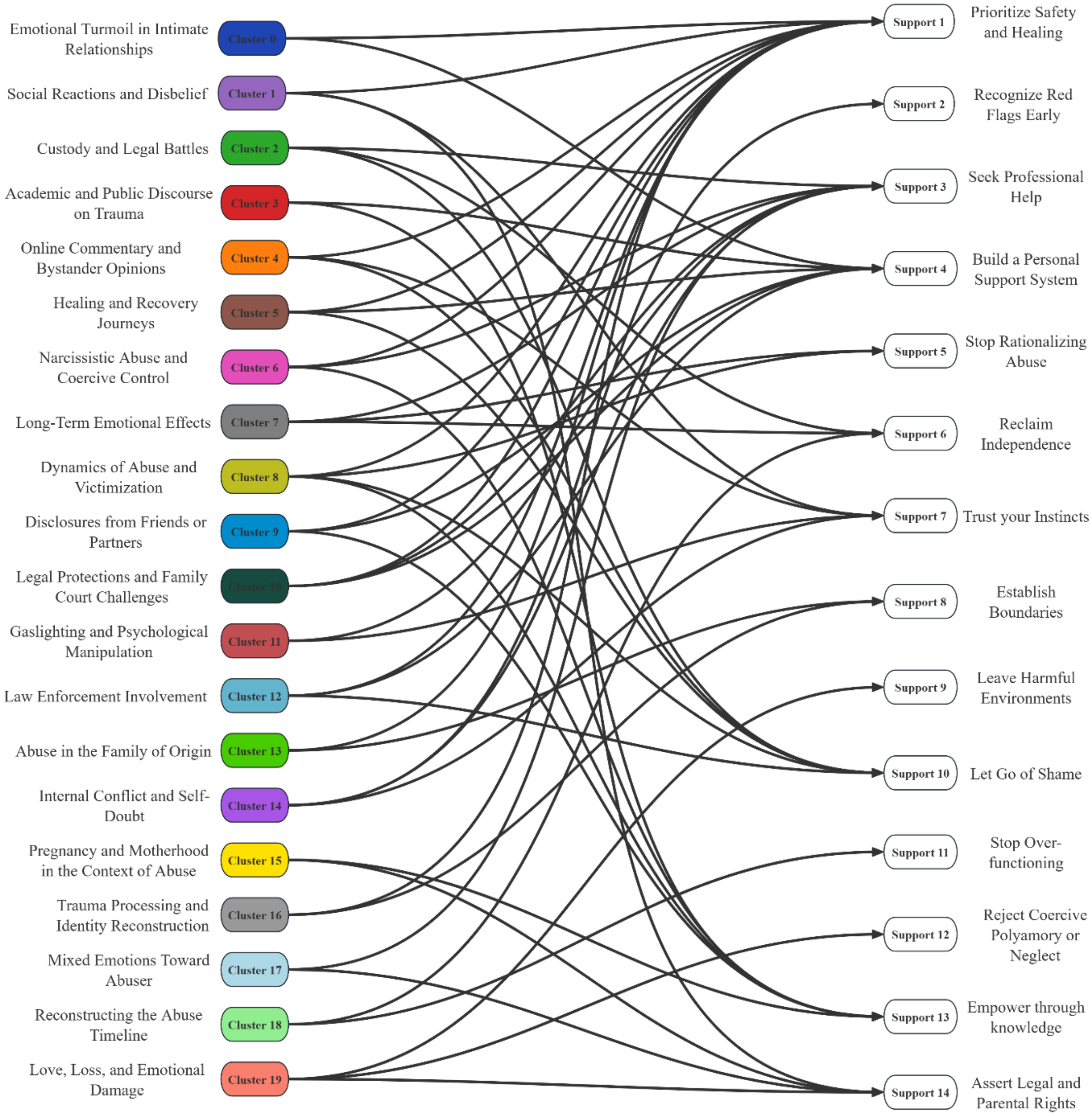

**Figure 4. DV self-disclosure topics and support provisions**

Other forms of support are more specific to certain types of abuse or emotional challenges. For example, Support 6 (reclaiming independence) helped in posts about custody battles, long-term effects, and piecing together abuse histories. Support 7 (trusting your instincts) is associated with clusters related to disbelief, gaslighting, and internal doubt, highlighting the importance of victims trusting their own feelings.

Support 5 (stop rationalizing abuse) and Support 8 (establish boundaries) are used to support posts about emotional control and self-blame. Less common but still important, Support 2 (recognize red flags early) addresses posts on internal conflict. Supports 9, 11, and 12 appear in deeply emotional posts about leaving harmful relationships, bearing excessive responsibility, or experiencing neglect, respectively.

## 6. Discussion

In addressing RQ1, we propose an integrated framework that leverages LLMs to effectively differentiate self-disclosure from non-self-disclosure content on social media platforms. Building on these results, we identified 20 distinct thematic areas of DV experiences from self-disclosure content in response to RQ2. Finally, we answered RQ3 by mapping these thematic areas to the categories of support mechanisms found within online community comments, such as emotional support, psychological support, and legal assistance.

Future studies should invite DV domain experts to investigate temporal patterns of online DV self-disclosure to better capture evolving victim support needs over extended periods. Additionally, the detection performance for self-disclosure is satisfactory but still has much room for improvement. Incorporating multimodal data, such as images and audio, could deepen our understanding of the thematic patterns of DV and enhance the performance of self-disclosure detection. Extending the current framework to additional social platforms and diverse languages would also allow for broader assessments of cultural applicability and platform-specific support mechanisms. Finally, incorporating these research insights into practical tools such as AI-driven chatbots or real-time digital support platforms could significantly enhance the accessibility and immediacy of support for DV victims seeking assistance online.

## Acknowledgment

This research was supported by the Fairfield University Summer Research Stipend, the Dolan School of Business Summer Research Grant, and the research grant from the School of Data Science at the University of North Carolina at Charlotte. The opinions, findings, conclusions, and recommendations presented in this article are those of the authors alone and do not necessarily represent the views of the funding organizations.